\documentclass[twocolumn]{aastex63}

\usepackage{wrapfig}
\usepackage{multirow}
\usepackage{lineno}

\received{-}
\revised{-}
\accepted{-}
\submitjournal{ApJL}

\shorttitle{C$_2$H$_3$NH$_2$ and C$_2$H$_5$NH$_2$ in the ISM}
\shortauthors{Zeng et al.}
\graphicspath{{./}{figures/}}

\begin{document}

\title{Probing the chemical complexity of amines in the ISM: detection of vinylamine (C$_2$H$_3$NH$_2$) and tentative detection of ethylamine (C$_2$H$_5$NH$_2$)}

\correspondingauthor{Shaoshan Zeng}
\email{shaoshan.zeng@riken.jp}

\author{Shaoshan Zeng}
\affiliation{Star and Planet Formation Laboratory, Cluster for Pioneering Research, RIKEN, 2-1 Hirosawa, Wako, Saitama, 351-0198, Japan}

\author{Izaskun Jim\'enez-Serra}
\affiliation{Centro de Astrobiolog\'ia (CSIC-INTA), Ctra. Ajalvir km 4, Torrej\'on de Ardoz, 28850, Madrid, Spain}

\author{V\'ictor M. Rivilla}
\affiliation{Centro de Astrobiolog\'ia (CSIC-INTA), Ctra. Ajalvir km 4, Torrej\'on de Ardoz, 28850, Madrid, Spain}
\affiliation{INAF-Osservatorio Astrofisico di Arcetri, Largo E. Fermi 5, I-50125, Florence, Italy}

\author{Jes\'us Mart\'in-Pintado}
\affiliation{Centro de Astrobiolog\'ia (CSIC-INTA), Ctra. Ajalvir km 4, Torrej\'on de Ardoz, 28850, Madrid, Spain}

\author{Lucas F. Rodr\'iguez-Almeida}
\affiliation{Centro de Astrobiolog\'ia (CSIC-INTA), Ctra. Ajalvir km 4, Torrej\'on de Ardoz, 28850, Madrid, Spain}

\author{Bel\'en Tercero}
\affiliation{Observatorio de Yebes (IGN), Cerro de la Palera s/n, 19141, Guadalajara, Spain}

\author{Pablo de Vicente}
\affiliation{Observatorio de Yebes (IGN), Cerro de la Palera s/n, 19141, Guadalajara, Spain}

\author{Fernando Rico-Villas}
\affiliation{Centro de Astrobiolog\'ia (CSIC-INTA), Ctra. Ajalvir km 4, Torrej\'on de Ardoz, 28850, Madrid, Spain}

\author{Laura Colzi}
\affiliation{Centro de Astrobiolog\'ia (CSIC-INTA), Ctra. Ajalvir km 4, Torrej\'on de Ardoz, 28850, Madrid, Spain}
\affiliation{INAF-Osservatorio Astrofisico di Arcetri, Largo E. Fermi 5, I-50125, Florence, Italy}

\author{Sergio Mart\'in}
\affiliation{Eureopean Southern Observatory, Alonso de C\'ordova 3107, Vitacura 763 0355, Santiago, Chile}
\affiliation{Joint ALMA Observatory, Alonso de C\'ordova 3107, Vitacura 763 0355, Santiago, Chile}

\author{Miguel A. Requena-Torres}
\affiliation{University of Maryland, College Park, ND 20742-2421, USA}
\affiliation{Department of Physics, Astronomy and Geosciences, Towson University, MD 21252, USA}



\begin{abstract}

Amines, in particular primary amines (R-NH$_2$) are closely related to the primordial synthesis of amino acids since they share the same structural backbone. However, only limited number of amines has been identified in the ISM which prevents us from studying their chemistry as well as their relation to pre-biotic species that could lead to the emergence of life. In this letter, we report the first interstellar detection of vinylamine (C$_2$H$_3$NH$_2$) and tentative detection of ethylamine (C$_2$H$_5$NH$_2$) towards the Galactic Centre cloud G+0.693-0.027. The derived abundance with respect to H$_2$ is (3.3$\pm$0.4)$\times$10$^{-10}$ and (1.9$\pm$0.5)$\times$10$^{-10}$, respectively. The inferred abundance ratios of C$_2$H$_3$NH$_2$ and C$_2$H$_5$NH$_2$ with respect to methylamine (CH$_3$NH$_2$) are $\sim$0.02 and $\sim$0.008 respectively. The derived abundance of C$_2$H$_3$NH$_2$, C$_2$H$_5$NH$_2$ and several other NH$_2$-bearing species are compared to those obtained towards high-mass and low-mass star-forming regions. Based on recent chemical and laboratory studies, possible chemical routes for the interstellar synthesis of C$_2$H$_3$NH$_2$ and C$_2$H$_5$NH$_2$ are discussed.

\end{abstract}

\keywords{Astrochemistry --- 
Chemical abundance --- Interstellar molecules --- Galactic Centre}


\section{Introduction}

The term prebiotic molecules refers to species that are considered to be involved in the processes leading to the origin of life. From the prospective of understanding the prebiotic chemistry in the interstellar medium (ISM), primary amines have caught much attention because they contain the same NH$_2$ group as those considered to be the fundamental building blocks of life (e.g. amino acids, nucleobases, nucleotides and other biochemical compounds). Although the attempts to observe glycine (NH$_2$CH$_2$COOH), the simplest amino acid, in the ISM have not succeeded \citep[e.g.][]{Ceccarelli2000,Belloche2013}, an increasing number of NH$_2$-bearing species have been reported in the last few years, which increases the chance to discover pre-biotic complex organics in space. This is in fact well-demonstrated towards G+0.693-0.027 (hereafter G+0.693), a molecular cloud located within the Sgr B2 complex in the Galactic Centre. As revealed by the census of N-bearing species towards G+0.693 \citep{Zeng2018}, several species containing NH$_2$ such as cyanamide (NH$_2$CN), formamide (NH$_2$CHO), and methylamine (CH$_3$NH$_2$) have been detected with abundance $\geq$10$^{-9}$. Following the search, two key precursors in the synthesis of prebiotic nucleotides, hydroxylamine (NH$_2$OH) and urea (NH$_2$CONH$_2$) have also been identified towards G+0.693 \citep{Rivilla2020v,Jimenez-Serra2020}. The very recent discovery of by far the most complex amine, ethanolamine (NH$_2$CH$_2$CH$_2$OH), the simplest head group of phospholipids in cell membrances, towards G+0.693 attests the potential of this source for finding more complex molecules of prebiotic relevance \citep{Rivilla2021}. This has thus prompted us to keep hunting for more amines in order to understand the chemical processes yielding related ingredients for life in space.

Structurally analogous to amino acids, CH$_3$NH$_2$ is the only primary amine that has been unambiguously observed towards different astronomical objects \citep[e.g.][]{Kaifu1974,Zeng2018,Bogelund2019,Ohishi2019}. Vinylamine (also known as ethenamine, C$_2$H$_3$NH$_2$) and ethylamine (C$_2$H$_5$NH$_2$) have not yet been reported in the ISM although the latter has been identified in multiple meteorites \citep[][and references therein]{Aponte2020} and materials returned by the \textit{Stardust} mission from comet 81P/Wild 2 \citep[e.g.][]{Glavin2008}. Due to some concerns on the possible contamination of the stardust sample, their cometary origin could not be confirmed until the robust detection in the coma of comet 67P/Churyumov-Gerasimenko \citep{Altwegg2016}. The detection of CH$_3$NH$_2$ and C$_2$H$_5$NH$_2$ together with glycine in these solar system objects enhanced the probability of amino acids being formed in space. Indeed, the retrosynthesis of amino acids revealed that molecules containing the -NH$_2$ functional group are likely precursors of amino acids \citep{Forstel2017}. For example, CH$_3$NH$_2$ and C$_2$H$_5$NH$_2$ may be the major constituent of glycine and alanine (NH$_2$CH$_3$CHCOOH) respecitvely. According to the detailed quantum chemical calculation, C$_2$H$_3$NH$_2$ is the next energetically stable isomer in C$_2$H$_5$N group after E- and Z- conformer of ethanimine (CH$_3$CHNH) \citep{Sil2018}, both of which have been detected in Sgr B2 \citep{Loomis2013} as well as in G+0.693 (Rivilla et al. in prep.). C$_2$H$_5$NH$_2$ exists in two forms: anti- and gauche-conformer. The former is known to be more stable and has the higher expected intensity ratio than the latter conformer \citep[see][for details]{Sil2018}. Therefore anti-C$_2$H$_5$NH$_2$ should be the most viable candidate for the astronomical detection in the C$_2$H$_7$N group. In this letter, we present the first detection of C$_2$H$_3$NH$_2$ and tentative detection of C$_2$H$_5$NH$_2$ in the ISM towards G+0.693 through the identification of several rotational transitions of its millimeter spectrum. 

\begin{deluxetable*}{ccccccccc}
  \tabletypesize{\scriptsize}
  \tablecolumns{9}
  \tablewidth{\textwidth}
  \tablecaption{List of observed transitions of C$_2$H$_3$NH$_2$ and C$_2$H$_5$NH$_2$. The following parameters are obtained from the CDMS catalogue entry 43504 and 45515: frequencies, quantum numbers, upper state degeneracy (g$_u$), the logarithm of the Einstein coefficients (log A$_{ul}$), and the energy of the upper levels (E$_u$). The derived root mean square (rms) of the anaylsed spectra region, integrated intensity ($\int T^*_A\,dv$), signal-to-noise ratio (S/N), and the information about the species with transitions slightly blended with C$_2$H$_3$NH$_2$ or C$_2$H$_5$NH$_2$ lines are provided in the last column. \label{tab:obs_transitions}}
  \tablehead{\colhead{Frequency} & \colhead{Transition} &  \colhead{log A$_{ul}$} & \colhead{g$_u$} & \colhead{E$_u$} & \colhead{rms} & \colhead{$\int T^*_A\,dv$} & \colhead{S/N$^{a}$} & \colhead{Blending} \\
  \colhead{(GHz)} & \colhead{(J$_{K_a,K_c}$)} &  \colhead{(s$^{-1}$)} & \colhead{} & \colhead{(K)} & \colhead{(mK)} & \colhead{(mK km s$^{-1}$)} & \colhead{} & \colhead{}}
  \startdata
    \multicolumn{9}{c}{C$_2$H$_3$NH$_2$} \\
    \hline
    92.31229 & 5$_{0,5}$-4$_{0,4}$,0$^+$ & -5.31611 & 11 & 13.3 & 1.2 & 325 & \multirow{2}{*}{46} & \multirow{2}{*}{aGg-(CH$_2$OH)$_2$}\\
    92.31539 & 5$_{0,5}$-4$_{0,4}$,0$^-$ & -5.41309 & 11 & 78.3 & 1.2 & 8 &  & \\
    $^*$92.92085 & 5$_{2,4}$-4$_{2,3}$,0$^+$ & -5.38264 & 11 & 22.4 & 1.3 & 170 & 22 & clean  \\
    $^*$96.51369 & 5$_{1,4}$-4$_{1,3}$,0$^+$ & -5.27533 & 11 & 16.1 & 2.5 & 282 & 19 & clean\\
    112.62479 & 6$_{2,4}$-5$_{2,3}$,0$^+$ & -5.10095 & 13 & 27.8 & 5.7 & 198 & 6 & g-C$_2$H$_5$SH \\
    $^*$128.32030 & 7$_{0,7}$-6$_{0,6}$,0$^+$ & -4.87614 & 15 & 24.7 & 8.7 & 350 & 7 & clean \\
    $^*$129.92445 & 7$_{2,6}$-6$_{2,5}$,0$^+$ & -4.89588 & 15 & 33.9 & 6.8 & 200 & 5 & clean \\
    $^*$146.03934 & 8$_{0,8}$-7$_{0,7}$,0$^+$ & -4.70420 & 17 & 31.7 & 2.9 & 316 & 20 & clean \\
    149.14290 & 8$_{3,6}$-7$_{3,5}$,0$^+$ & -4.74122 & 17 & 52.4 & 2.9 & 93 & 6 & CH$_3$C$^{13}$CH \\
    159.88701 & 9$_{1,9}$-8$_{1,8}$,0$^+$ & -4.58763 & 19 & 40.7 & 5.3 & 241 & 9 & C$_2$H$_5$CN \\
    \hline
    \multicolumn{9}{c}{C$_2$H$_5$NH$_2$} \\
    \hline
    $^*$32.14604 & 2$_{1,2}$-1$_{1,1}$,0$^-$ & -6.88737 & 15 & 3.4 & 1.1 & 8 & \multirow{2}{*}{4} & \multirow{2}{*}{clean} \\
   $^*$ 32.14605 & 2$_{1,2}$-1$_{1,1}$,0$^+$ & -6.88739 & 45 & 3.4 & 1.1 & 24 & & \\
    $^*$34.04659 & 2$_{1,1}$-1$_{1,0}$,0$^-$ & -6.81259 & 15 & 3.5 & 1.2 & 8 & \multirow{2}{*}{4} & \multirow{2}{*}{clean} \\
    $^*$34.04662 & 2$_{1,1}$-1$_{1,0}$,0$^+$ & -6.81261 & 45 & 3.5 & 1.2 & 25 & &  \\
    $^*$49.52875 & 3$_{0,3}$-2$_{0,2}$,0$^+$ & -6.16968 & 63 & 4.7 & 2.8 & 67 & \multirow{2}{*}{6} & \multirow{2}{*}{clean} \\
    $^*$49.52875 & 3$_{0,3}$-2$_{0,2}$,0$^-$ & -6.16966 & 21 & 4.7 & 2.8 & 22 & &  \\
    82.16878 & 5$_{0,5}$-4$_{0,4}$,0$^+$ & -5.48528 & 99 & 11.8 & 2.8 & 109 & \multirow{2}{*}{8} & \multirow{2}{*}{$^{13}$CH$_3$CH$_2$OH} \\
    82.16878 & 5$_{0,5}$-4$_{0,4}$,0$^-$ & -5.48526 & 33 & 11.8 & 2.8 & 36 & &  \\
   $^*$ 83.24287 & 5$_{2,3}$-4$_{2,2}$,0$^+$ & -5.54312 &99 & 16.4 & 3.1 & 64 & \multirow{2}{*}{4} & \multirow{2}{*}{clean} \\ 
   $^*$83.24287 & 5$_{2,3}$-4$_{2,2}$,0$^-$ & -5.54310 & 33 & 16.4 & 3.1 & 21 & &  \\
    84.98078 & 5$_{1,4}$-4$_{1,3}$,0$^+$ & -5.45828 & 99 & 13.3 & 3.3 & 96 & \multirow{2}{*}{6} & \multirow{2}{*}{H$^{15}$NCO} \\
    84.98078 & 5$_{1,4}$-4$_{1,3}$,0$^-$ & -5.45826 & 33 & 13.3 & 3.3 & 32 & & \\
    145.87174 & 9$_{0,9}$-8$_{0,8}$,0$^+$ & -4.72105 & 171 & 35.3 & 2.3 & 56 & \multirow{2}{*}{6} & \multirow{2}{*}{S$^{18}$O} \\
    145.87174 & 9$_{0,9}$-8$_{0,8}$,0$^-$ & -4.72103 & 57 & 35.3 & 2.3 & 19 & & \\
  \enddata
  \tablenotetext{a}{S/N is calculated as: $(\int T_{A}^{*}dv)/[\rm rms(\frac{\Delta \nu}{\rm FWHM})^{0.5}\rm FWHM]$, where $\Delta \nu$ is the spectral resolution of the data, range between 1.5 and 2.2 km s$^{-1}$. The clean transitions are denoted by an asterisk. A single common S/N is given for the overlapping transitions.}
\end{deluxetable*}

\begin{figure*}
\centering
   \includegraphics[width=0.8\textwidth]{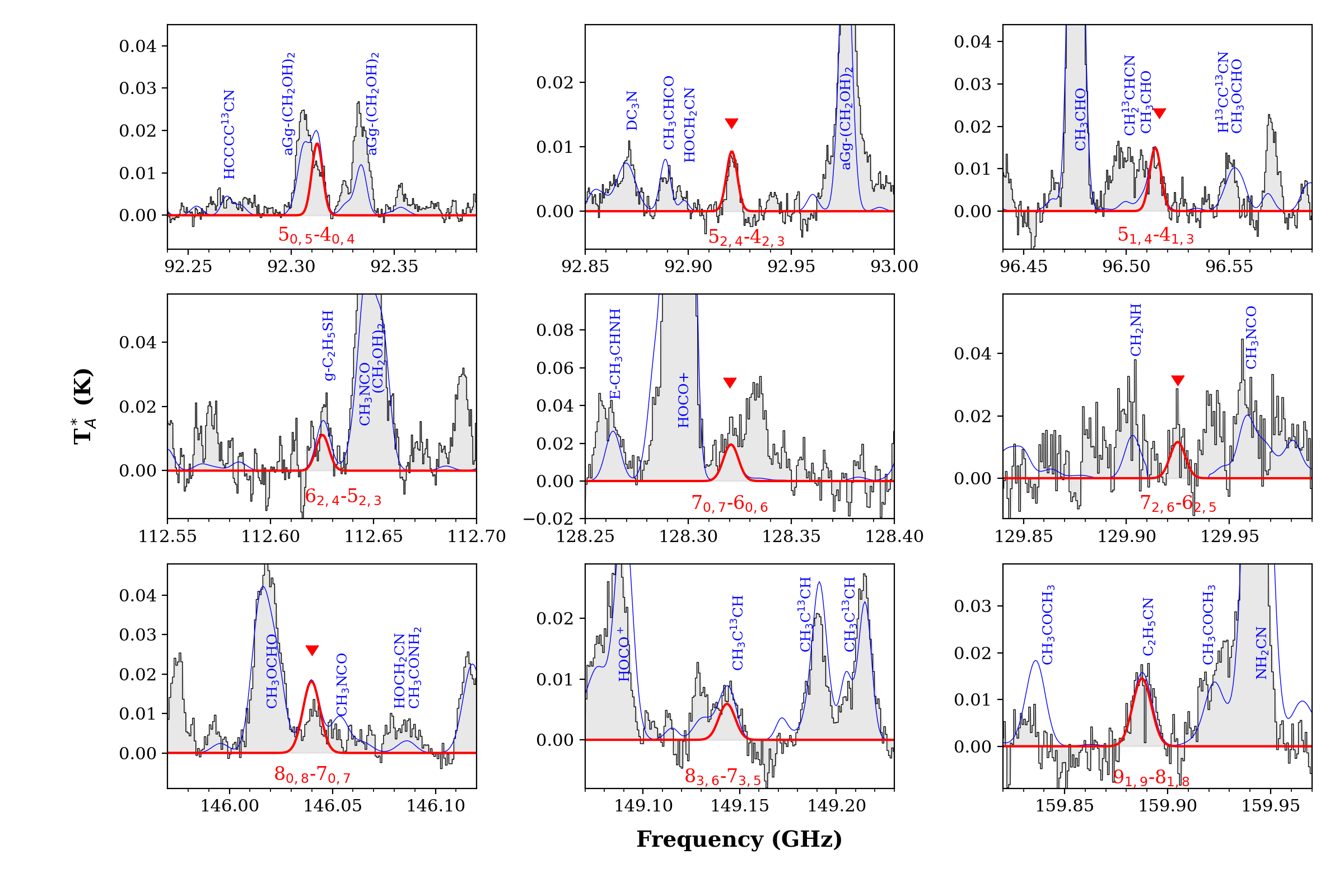}  
   \vspace{-0.5cm}
\caption{Unblended or only slightly blended transitions of C$_2$H$_3$NH$_2$ detected towards G+0.693. The red line shows the best LTE fit to the C$_2$H$_3$NH$_2$ lines while the blue line shows the total contribution, including the emission from other molecular species (labelled) identified in G+0.693. The cleanest detected transitions are denoted by a red $\blacktriangledown$.}
 \label{fig:mol_spectra_1}
\end{figure*}
\begin{figure*}
\centering
    \includegraphics[width=\textwidth]{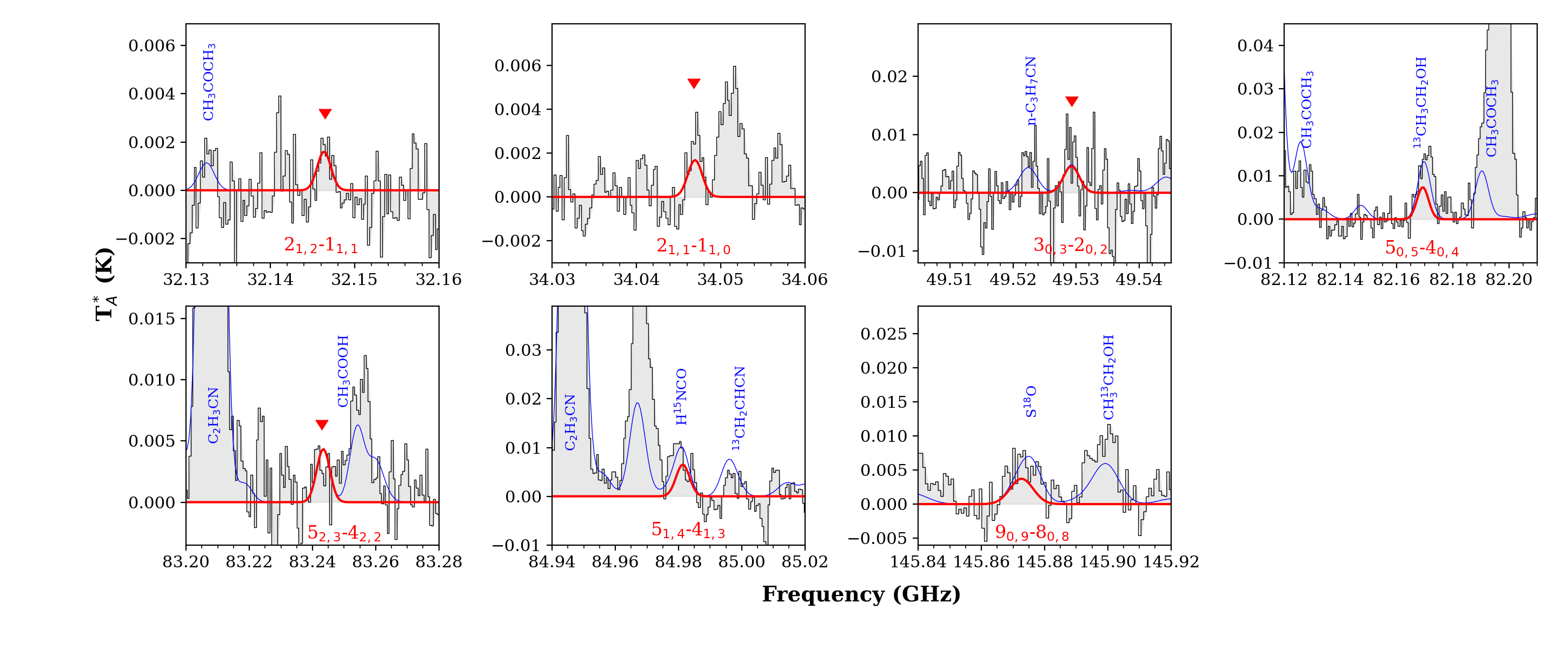}
       \vspace{-0.8cm}
\caption{Unblended or only slightly blended transitions tentatively identified for C$_2$H$_5$NH$_2$ towards G+0.693. The cleanest transitions are denoted by a red $\blacktriangledown$.}
 \label{fig:mol_spectra_2}
\end{figure*}

\section{Observations} \label{obs}
We have carried out high-sensitivity spectral surveys at 7, 3, and 2$\,$mm towards G+0.693 molecular cloud using the IRAM 30m\footnote{IRAM is supported by INSU/CNRS (France), MPG (Germany), and IGN (Spain)} and Yebes 40m\footnote{http://rt40m.oan.es/rt40m en.php} telescopes. The observations were centred at $\alpha$(J2000) = 17$^h$47$^m$22$^s$, $\delta$(J2000) = -28$^{\circ}$21$'$27$''$. The position switching mode was used in all observations with the reference position located at $\Delta\alpha$, $\Delta\delta$ = -885$^{''}$,290${''}$ with respect to the source position. The half-power beam width (HPBW) of the IRAM 30m and Yebes 40m telescope are in a range of 14$^{\prime\prime}$-36$^{\prime\prime}$ at observed frequencies between 30 GHz and 175 GHz. The intensity of the spectra was measured in unit of antenna temperature, T$_{\rm A}^*$ as the molecular emission toward G+0.693 is extended over the beam \citep{Zeng2020}. The IRAM 30$\,$m observations were performed in three observing runs during 2019: April 10-16, August 13-19 and December 11-15, from projects numbers 172-18 (PI Mart\'in-Pintado), 018-19 (PI Rivilla) and 133-19 (PI Rivilla). It covered spectral ranges of 71.76-116.72$\,$GHz and 124.77-175.5$\,$GHz. We refer to \citet{Rivilla2020v} for a full description of the IRAM 30$\,$m observations. The Yebes 40$\,$m observations were carried out during 6 observing sessions in February 2020, as part of the project 20A008 (PI Jim\'enez-Serra). The new Q band (7mm) HEMT receiver was used to allow broad-band observations in two linear polarisations. The spectral coverage ranges from 31.075 GHz to 50.424 GHz. We refer to \citet{Zeng2020} and \citet{Rivilla2020v} for more detailed information on the Yebes 40$\,$m observations.

\section{Analysis and Results} \label{ana}
The line identification and analysis were carried out using the SLIM (Spectral Line Identification and Modelling) tool implemented within the \textsc{madcuba} package\footnote{Madrid Data Cube Analysis on ImageJ is a software developed at the Center of Astrobiology (CAB) in Madrid; http://cab.inta-csic.es/madcuba/Portada.html.}\citep[version 21/12/2020,][]{Martin2019}. The spectroscopic information of C$_2$H$_3$NH$_2$, within 0$^+ \&$ 0$^-$ (CDMS\footnote{Cologne Database for Molecular Spectroscopy \citep{Endres2016}. https://cdms.astro.uni-koeln.de/classic/} entry 43504) was obtained from \citet{Brown1990} and \citet{Mcnaughton1994}. And the spectroscopic information of C$_2$H$_5$NH$_2$, anti-conformer (CDMS entry 45515) was obtained from \citet{Fischer1982} and \citet{Apponi2008}. Table \ref{tab:obs_transitions} summarises the unblended or only partially blended transitions of C$_2$H$_3$NH$_2$ and C$_2$H$_5$NH$_2$ detected towards G+0.693. Note that all the C$_2$H$_5$NH$_2$ lines in Table \ref{tab:obs_transitions} are a-type transitions with selection rules of $\Delta$K$_a$=0 and $\Delta$K$_c$=$\pm$1. The rotational spectrum of C$_2$H$_3$NH$_2$ and C$_2$H$_5$NH$_2$ are characterised by inversion doubling due to the large amplitude inversion motion of the NH$_2$ group. Each rotational energy level, specified by the rotational quantum numbers J and K, is thus split into an inversion doublet. The 0$^+$ or 0$^-$ in Table \ref{tab:obs_transitions} indicate the inversion doublet from which the transition arises. For C$_2$H$_5$NH$_2$, the variation of upper state degeneracy and integrated intensity of the same transition is due to the spin statistical weight of 3:1 between the symmetric and anti-symmetric sub-levels.

The analysis was performed under the assumption of local thermodynamic equilibrium (LTE) conditions due to the lack of collisional coefficients of C$_2$H$_3$NH$_2$ and C$_2$H$_5$NH$_2$. Due to the low density of G+0.693 \citep[$\sim$10$^4$-10$^5$ cm$^{-3}$;][]{Zeng2020}, molecules are sub-thermally excited in the source and hence their excitation temperatures \citep[in a range of 5 to 20 K; e.g.][]{Requena-torres2008,Rivilla2018,Zeng2018} are significantly lower than the kinetic temperature of the source \citep[$\sim$150 K; e.g.][]{Zeng2018}. However, note that the transitions of C$_2$H$_3$NH$_2$ and C$_2$H$_5$NH$_2$ are well fitted using one excitation temperature for each species. Considering the effect of line opacity, MADCUBA-SLIM generated synthetic spectra that can be compared to the observed spectra. The MADCUBA-AUTOFIT tool was then used to provide the best non-linear least-squares LTE fit to the data using the Levenberg-Marquardt algorithm. It is important to note that not a single transition of C$_2$H$_3$NH$_2$ and C$_2$H$_5$NH$_2$ predicted by the LTE spectrum is missing in the data. To properly evaluate the line contamination by other molecules, over 300 species have been searched for in our dataset. This included not only all the molecules detected towards G+0.693 in previous studies \citep{Requena-torres2008, Zeng2018, Rivilla2018, Rivilla2019, Rivilla2020v, Rivilla2021, Jimenez-Serra2020, Rodriguez-Almeida2021}, but also those reported in the ISM\footnote{See https://cdms.astro.uni-koeln.de/classic/molecules}. The molecule blended with the transition of C$_2$H$_3$NH$_2$ and C$_2$H$_5$NH$_2$ is listed in the last column of Table \ref{tab:obs_transitions}. We note that the line identification and fitting file used in this analysis is the same as the one used in previous works \citep[e.g.][]{Zeng2018,Rodriguez-Almeida2021} as well as on-going works. Therefore, the excitation temperatures and column densities of blending species are consistent with the ones reported in all these studies.

For C$_2$H$_3$NH$_2$, we detect five clean transitions and four slightly blended transitions with a blending contribution of $<$10$\%$ (see Figure \ref{fig:mol_spectra_1}). For C$_2$H$_5$NH$_2$, only four clean transitions are reported and three are slightly blended (Figure \ref{fig:mol_spectra_2}). Since the clean transitions of C$_2$H$_5$NH$_2$ are weak (S/N=4 in integrated intensity; Table \ref{tab:obs_transitions}), we conclude that this species is tentatively detected.
The free parameters that can be fitted are: molecular column density ($\textit{N}_{\rm tot}$), excitation temperature ($\textit{T}_{\rm ex}$), radial velocity ($\textit{V}_{\rm LSR}$), full width half maximum (\textit{FWHM}, $\Delta \nu$), and source size ($\theta$). For G+0.693, we assumed that the source size is extended in \textsc{madcuba}. As the algorithm did not converge when fitting C$_2$H$_3$NH$_2$ with all parameters left free, we fixed the \textit{FWHM} and $\textit{V}_{\rm LSR}$ to 18$\,$km s$^{-1}$ and 67 $\,$km s$^{-1}$ respectively by visual inspection of the most unblended lines. These values are consistent with those from many other molecules previously analysed in G+0.693 \citep[\textit{FWHM}$\sim$20 km s$^{-1}$, $\textit{V}_{\rm LSR} \sim$68$\,$km s$^{-1}$;][]{Requena-torres2008, Zeng2018, Rivilla2018, Rivilla2019, Rivilla2020v, Jimenez-Serra2020, Rodriguez-Almeida2021}. The resulting LTE fit gives $\textit{T}_{\rm ex}$=(18$\pm$3) K and $\textit{N}_{\rm tot}$=(4.5$\pm$0.6)$\times$10$^{13}$ cm$^{-2}$. In the case of C$_2$H$_5$NH$_2$, the $\textit{V}_{\rm LSR}$ was fixed to 67$\,$km s$^{-1}$ and the LTE fit gives $\textit{T}_{\rm ex}$=(12$\pm$5) K, \textit{FWHM}=(18$\pm$5)$\,$km s$^{-1}$ and $\textit{N}_{\rm tot}$=(2.5$\pm$0.7)$\times$10$^{13}$ cm$^{-2}$.


The resulting best LTE fit to the C$_2$H$_3$NH$_2$ and C$_2$H$_5$NH$_2$ lines is shown in red line in Fig.\ref{fig:mol_spectra_1} and Fig.\ref{fig:mol_spectra_2} whilst blue line indicates the best fit considering also the total contribution of the LTE emission from all the other identified molecules. Adapting the H$_2$ column density inferred from observations of C$^{18}$O \citep[$\textit{N}_{\rm H_2}$ = 1.35$\times$10$^{23}$ cm$^{-2}$;][]{Martin2008}, the resulting abundance is (3.3$\pm$0.4)$\times$10$^{-10}$ and (1.9$\pm$0.5)$\times$10$^{-10}$ for C$_2$H$_3$NH$_2$ and C$_2$H$_5$NH$_2$ respectively.

\section{Discussion} \label{dis}

\begin{figure*}
\vspace{-0.3 cm}
\centering
   \includegraphics[width=\textwidth]{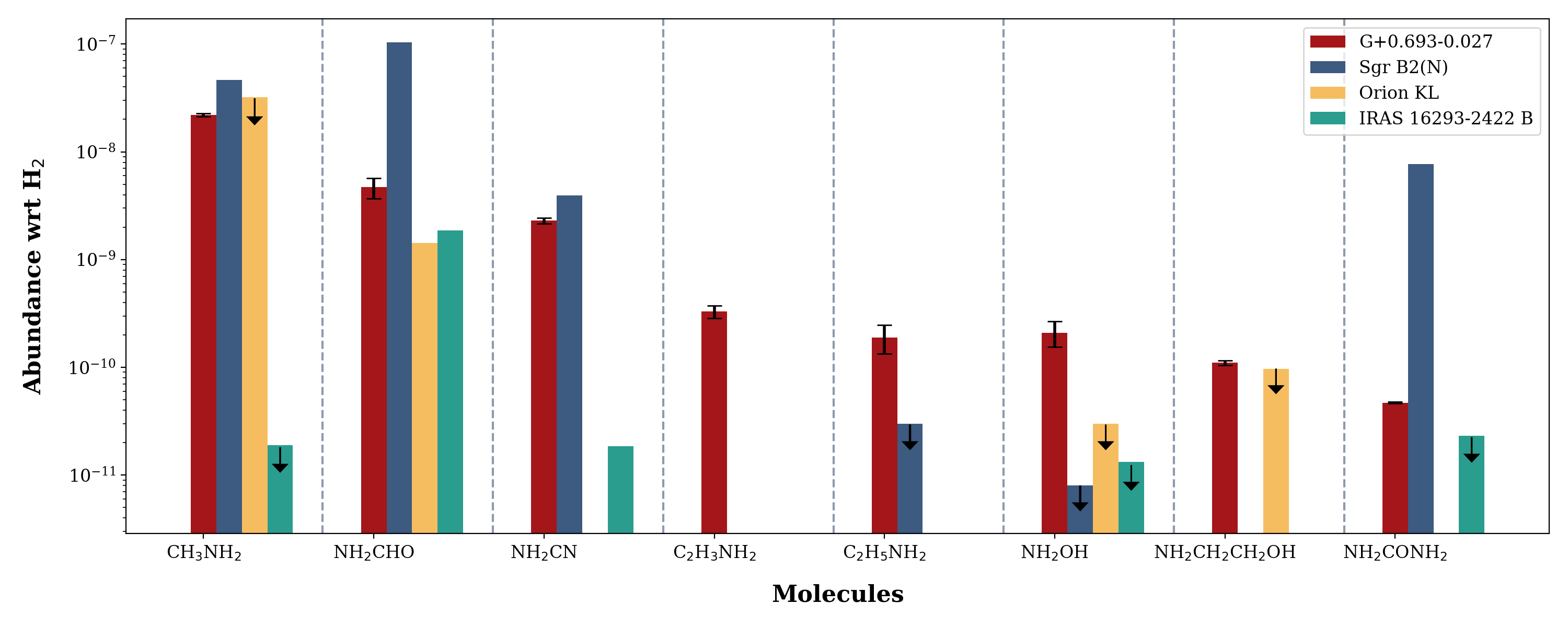}  
\caption{Derived abundance of NH$_2$-bearing species with respect to H$_2$ towards G+0.693-0.027, Sgr B2(N), Orion KL, and IRAS 16293-2422 B. For G+0.693-0.027 \citep[$N(\rm H_2$)=1.35$\times$10$^{23}$ cm$^{-2}$;][]{Martin2008}, molecular column densities are obtained from this work, \citet{Zeng2018,Jimenez-Serra2020, Rivilla2020v} and \citet{Rivilla2021}. For IRAS 16293-2422 B \citep[$N(\rm H_2$)=2.8$\times$10$^{25}$ cm$^{-2}$;][]{Martin-Domenech2017}, molecular column densities are from \citet{Martin-Domenech2017,Ligterink2018} and \citet{Jimenez-Serra2020}. For Sgr B2(N): CH$_3$NH$_2$, NH$_2$CHO, and NH$_2$CN from \citet{Belloche2013} with $N(\rm H_2$)=1.3$\times$10$^{25}$ cm$^{-2}$ \citep{Belloche2008}; abundance of C$_2$H$_5$NH$_2$ and upper limit of NH$_2$OH is taken directly from \citet{Apponi2008} and \citet{Pulliam2012} respectively; NH$_2$CONH$_2$ is derived from \citet{Belloche2019} by assuming Sgr B2(N1S) has the same $N(\rm H_2$)=3.5$\times$10$^{24}$ cm$^{-2}$ \citep{Li2021} as Sgr B2 (N1E). For Orion KL: column density of CH$_3$NH$_2$ and $N(\rm H_2$)=3.1$\times$10$^{23}$ cm$^{-2}$ are from \citet{Pagani2017}; NH$_2$CHO from \citet{Motiyenko2012} with $N(\rm H_2$)=4.2$\times$10$^{23}$ cm$^{-2}$ \cite{Tercero2010}; upper limits of NH$_2$OH and NH$_2$CH$_2$CH$_2$OH are from \citet{Pulliam2012} and \citet{Wirstrom2007} respectively with $N(\rm H_2$)=7.0$\times$10$^{23}$ cm$^{-2}$ \citep{Womack1992}.}
 \label{fig:comparison}
\end{figure*}

Figure \ref{fig:comparison} presents the abundance with respect to H$_2$, in decreasing order, of NH$_2$-bearing species detected towards G+0.693. The results are compared to those derived towards three high-mass and low-mass star-forming regions that are chemically rich, i.e. Sgr B2(N), Orion KL, and IRAS 16293-2422 B. The low abundance or non-detection of -NH$_2$ species may indicate that their formation is less efficient towards Orion KL and IRAS 16293-2422 B. On the other hand, the detection with abundance $>$10$^{-11}$ suggests G+0.693 is a prominent -NH$_2$ molecule repository which allows us to study their origin as well as their chemical relation to other pre-biotic molecules. 

\begin{figure}
\centering
   \includegraphics[width=\linewidth]{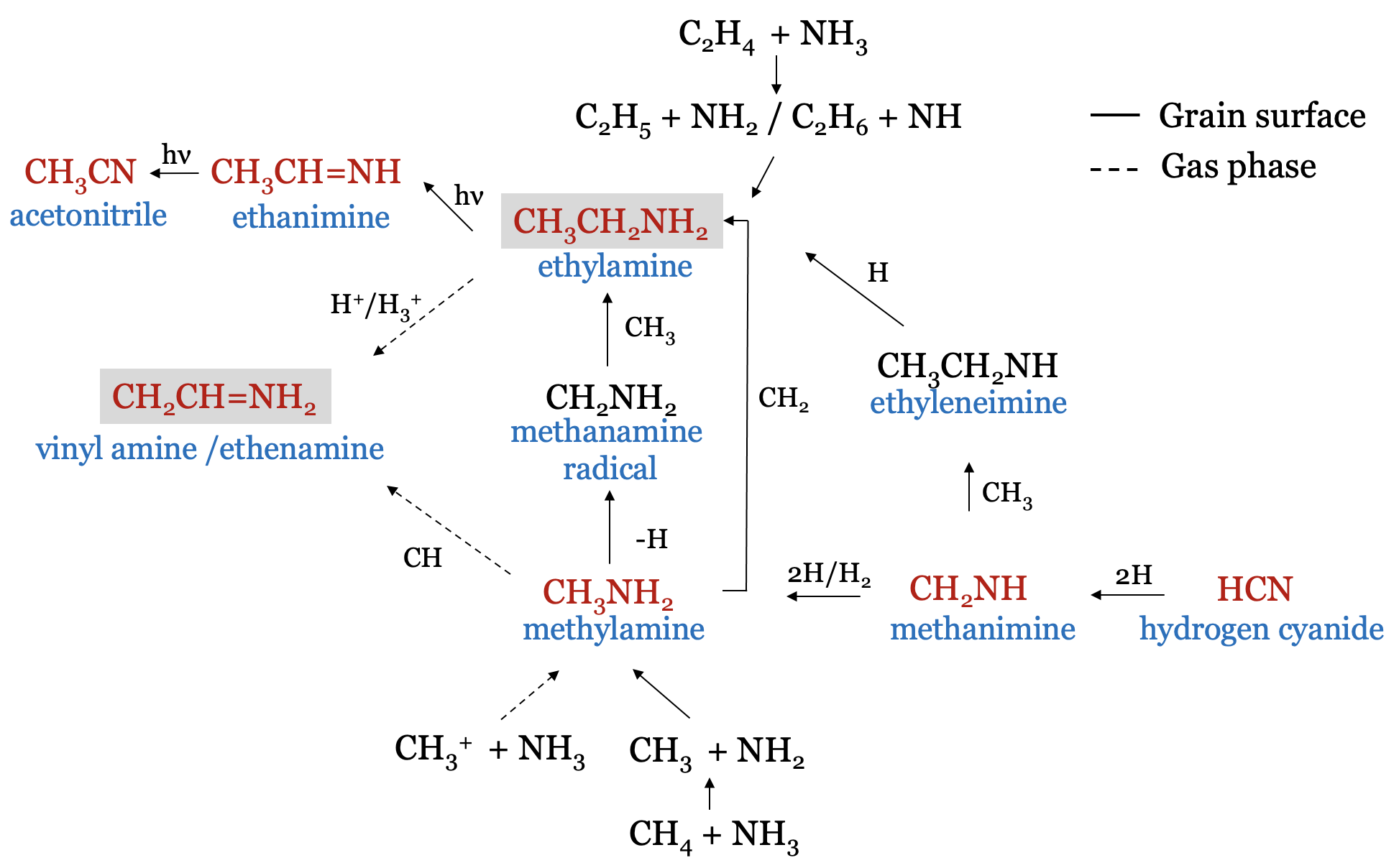}  
\caption{Summary of the chemical routes proposed for the formation of C$_2$H$_3$NH$_2$ and C$_2$H$_5$NH$_2$ in the ISM. The molecular species in red are those that have been detected towards G+0.693. The solid arrows denote surface chemistry reactions and dashed arrows denote gas-phase chemistry.}
 \label{fig:formation}
\end{figure}

In contrast to the three compared sources, G+0.693 lacks an internal heating source responsible for the rich chemistry. The high level of molecular complexity is attributed to dust grain sputtering by low-velocity shocks ($\leq$20 km s$^{-1}$), which is driven by the possible cloud-cloud collision occurring in the Sgr B2 complex \citep{Zeng2020}. This is indicated by the high abundances of shock tracers such as HNCO and SiO \citep{Martin2008,Rivilla2018} and the presence of molecules that are known to be formed on grain surfaces \citep{Requena-torres2008,Zeng2018}. In addition, the abundance ratio of HC$_3$N/HC$_5$N and C$_2$H$_3$CN/C$_2$H$_5$CN derived in \citet{Zeng2018} suggested that an enhanced cosmic-ray ionisation rate may also play a role in the chemistry of N-bearing species towards G+0.693. In the following Section, we evaluate the possible formation routes for the amines detected towards G+0.693 and the discussed formation routes are summarised in Figure \ref{fig:formation}. 


\subsection{Formation of primary amines}
Despite the low number of detections in the ISM, the CH$_3$NH$_2$ chemistry under astrophysical conditions has been studied in theoretical, experimental and chemical modelling work. In the gas-phase, CH$_3$NH$_2$ is proposed to form via the radiative association between ammonia (NH$_3$) and the methyl radical cation (CH$_3^+$) followed by recombination dissociation \citep{Herbst1985}. On the grain surfaces, experimental work demonstrated that CH$_3$NH$_2$ can form via sequential hydrogenation of hydrogen cyanide (HCN): HCN $\rightarrow$ CH$_2$NH $\rightarrow$ CH$_3$NH$_2$ \citep{Theule2011}. 
Alternatively, gas-grain chemical model by \citet{Garrod2008} has suggested that CH$_3$NH$_2$ is formed by simple addition of CH$_3$ from CH$_4$ to azanyl radical (NH$_2$) from NH$_3$ during warm-up phases. This radical-radical recombination has recently been studied in laboratory ice simulations revealing that CH$_3$NH$_2$ can be synthesised in irradiated ices composed of CH$_4$ and NH$_3$ \citep{Kim2011, Forstel2017} but also in cold and quiescent molecular clouds \citep{Ioppolo2021}. On the contrary, little is known about the chemistry of C$_2$H$_3$NH$_2$ and C$_2$H$_5$NH$_2$.

\textit{\textbf {C$_2$H$_5$NH$_2$:}} Although there are no chemical routes included in current astrochemical databases, different formation mechanisms of C$_2$H$_5$NH$_2$ have been discussed in the literature. For instance, the addition of the CH$_3$ radical to CH$_2$NH followed by hydrogenation of the resulting product would produce C$_2$H$_5$NH$_2$ on grain surfaces \citep{Bernstein1995}; the photochemistry of a mixture of ethylene (C$_2$H$_4$) and NH$_3$ in which C$_2$H$_5$NH$_2$ is formed from radical-radical reaction of the ethyl radical (C$_2$H$_5$) with NH$_2$ \citep[][and references therein]{Danger2011}. Reactions of CH$_3$NH$_2$ with carbene (CH$_2$) or ethane (C$_2$H$_6$) with nitrene (NH) are also expected to form C$_2$H$_5$NH$_2$, but with lower probability \citep{Forstel2017}. More recently, experimental simulation of interstellar ice analogs containing CH$_3$NH$_2$ revealed that C$_2$H$_5$NH$_2$ can be formed efficiently from CH$_3$ and methanamine radical (CH$_2$NH$_2$) \citep{Carrascosa2021}. 

In addition to the aforementioned formation pathways, one might also expect C$_2$H$_5$NH$_2$ to be formed through the same mechanism as suggested for CH$_3$NH$_2$: successive hydrogenation starting from acetonitrile (CH$_3$CN): CH$_3$CN $\rightarrow$ CH$_3$CHNH $\rightarrow$ C$_2$H$_5$NH$_2$. However, CH$_3$CN has been proposed to be a product of the surface chemistry of CH$_3$NH$_2$ \citep{Carrascosa2021}. Furthermore, laboratory work showed that CH$_3$CN does not react with H atoms between 10 and 60 K \citep{Nguyen2019}. Considering the temperature of dust grains in G+0.693 is T$_{\rm dust} \leq$30 K \citep[][and reference therein]{Zeng2018}, this proposed hydrogenation leading to C$_2$H$_5$NH$_2$ is unlikely to occur on dust grains in G+0.693. Based on the reaction rate coefficients provided in \citet{Sil2018}, C$_2$H$_5$NH$_2$ is less likely produced in the same hydrogenation reaction in the gas phase due to its low efficiency at typical kinetic temperature in Galactic Centre clouds, \textit{T}$_{\rm kin}$ = 50$-$120 K \citep[][and reference therein]{Zeng2018}. To our best knowledge, with no other possible gas phase reaction to form C$_2$H$_5$NH$_2$, this species likely forms on the surface of dust grains in G+0.693. Regardless the chemical formation route on grains, the sputtering of grain icy mantles by large-scale low-velocity ($\leq$20 km s$^{-1}$) shocks in G+0.693 would release C$_2$H$_5$NH$_2$ into gas phase from grains \citep[see e.g.][]{Requena-torres2008,Martin2008,Zeng2020}.

\textit{\textbf {C$_2$H$_3$NH$_2$:}} With the available reaction rate coefficients in the Kinetic Database for Astrochemistry \citep[KIDA,][]{Wakelam2012}, C$_2$H$_3$NH$_2$ is proposed to form from the reaction between the CH radical and CH$_3$NH$_2$ in the gas-phase, most efficient at temperatures T=50$-$200 K. With \textit{T}$_{\rm kin}$ = 50$-$120 K, C$_2$H$_3$NH$_2$ is thus expected to be formed efficiently via this pathway in G+0.693. In particular, CH$_3$NH$_2$ is found to be abundant \citep[$\sim$10$^{-8}$;][]{Zeng2018} in G+0.693 which would be readily available for this chemical reaction to proceed. Another possible gas-phase formation route is through the reaction involving C$_2$H$_5$NH$_2$ + H$^+$ or H$_3^+$ followed by recombination dissociation, analogous to the formation of C$_2$H$_3$CN from C$_2$H$_5$CN proposed by \citet{Caselli1993}. As discussed in \citet{Zeng2018} for -CN group species, this ion-molecule gas-phase reaction can be efficient thanks to (i) the presence of high cosmic ray ionisation rate in the Galactic Centre and (ii) the relatively low densities of $\sim$10$^4$ cm$^{-3}$ of G+0.693. This, and the fact that a similar abundance ratios are found for C$_2$H$_3$CN/C$_2$H$_5$CN = 2.2$\pm$0.3 and C$_2$H$_3$NH$_2$/C$_2$H$_5$NH$_2$=1.7$\pm$0.5 towards G+0.693, makes this formation route plausible.

On the grain surface, C$_2$H$_3$NH$_2$ might be a photoproduct of C$_2$H$_5$NH$_2$ \citep{Hamada1984}. But recent experimental investigation showed that its isomer, ethanimine (CH$_3$CHNH), appears to be the primary product of photolysis of C$_2$H$_5$NH$_2$ \citep{Danger2011}. This has also been recently found by \citet{Carrascosa2021}, who even synthesised large N-heterocycles in interstellar ice analogs under UV radiation. 
We thus propose that the formation of C$_2$H$_3$NH$_2$ likely occurs in the gas phase towards G+0.693 although further theoretical or laboratory work is needed to determine the rate constant of the reaction C$_2$H$_5$NH$_2$ + H$^+$/H$_3^+$.

In summary, we report the discovery of two new amines in the ISM: C$_2$H$_3$NH$_2$ and tentatively C$_2$H$_5$NH$_2$. The abundance ratios with respect to CH$_3$NH$_2$ are $\sim$0.02 and $\sim$0.008, i.e. about a factor $>$10. This trend has been found for other species such as ethanol (with respect to methanol) or ethyl cyanide \citep[with respect to CH$_3$CN;][]{Zeng2018,Rodriguez-Almeida2021}. Primary amines are known to be involved in the synthesis of proteinogenic alpha-amino acids \cite{Forstel2017}. Therefore, their discovery provides crucial information about the connection between interstellar chemistry and the prebiotic material found in meteorites and comets.

\acknowledgments

The authors wish to thank the referees for constructive comments that significantly improved the paper. We are grateful to the IRAM 30-m and Yebes 40-m telescope staff for help during the different observing runs. IRAM is supported by the National Institute for Universe Sciences and Astronomy/National Center for Scientific Research (France), Max Planck Society for the Advance- ment of Science (Germany), and the National Geographic Institute (IGN) (Spain). The 40-m radio telescope at Yebes Observatory is operated by the IGN, Ministerio de Transportes, Movilidad y Agenda Urbana. This study is supported by a Grant-in-Aid from the Ministry of Education, Culture, Sports, Science, and Technology of Japan (20H05845) and by a RIKEN pioneering Project (Evolution of Matter in the Universe). We also acknowledge partial support from the Spanish National Research Council (CSIC) through the i-Link project number LINKA20353.  I.J.-S. and J.M.-P. have received partial support from the Spanish State Research Agency through project number PID2019-105552RB-C41. V.M.R., L.R.-A., and L.C. have received funding from the Comunidad de Madrid through the Atracci\'on de Talento Investigador (Doctores con experiencia) Grant (COOL: Cosmic Origins Of Life; 2019-T1/TIC-15379).
%







\bibliographystyle{aasjournal}
\bibliography{references}{}



\end{document}